\documentclass[aps,prl,twocolumn,showpacs,amsmath,amssymb]{revtex4-1}

\usepackage[dvips]{graphicx}
\usepackage{dcolumn}
\usepackage{bm}
\usepackage{color}
\def\be{\begin{equation}}
\def\en{\end{equation}}
\def\bea{\begin{eqnarray}}
\def\ena{\end{eqnarray}}
\def\bec{\begin{equation}\begin{array}{rcl}}

\def\p{\partial}
\def\ep{\epsilon}
\def\gs{\gtrsim}
\def\ls{\lesssim}

\def\ve{\varepsilon}
\newcommand{\av}[1]{\langle{#1}\rangle}

\newcommand{\bi}[1]{\mbox{\boldmath$#1$}}
\newcommand{\pp}[2]{\frac{\partial {#1}}{\partial {#2}}}

\def\rij{{\bi r}_{ij}}
\def\hrij{\hat{\bi r}_{ij}}

\def\aQ{\stackrel{\leftrightarrow}{Q}}
\def\aI{\stackrel{\leftrightarrow}{I}}

\begin{document}
\title{
Structural phase transition 
of anisotropic particles 
and formation of  orientation-strain glass 
with addition of impurities 
}  
\author{Kyohei Takae and Akira Onuki}
\affiliation{Department of Physics, Kyoto University, Kyoto 606-8502, Japan}


\date{\today}

\begin{abstract} 
Using  a modified Lennard-Jones 
model for anisotropic particles, we 
present results of molecular dynamics simulation 
in two dimensions.   
In one-component systems,  we find   
crystallization,  
 a Berezinskii-Kosterlitz-Thouless  phase,  
and  a structural phase  transition,  
as the  temperatures 
is lowered. In the
lowest temperature range, 
the crystal is composed of three martensitic 
variants on a hexagonal lattice, exhibiting  the shape memory effect.  
With addition of larger spherical 
particles (impurities), 
these domains are finely divided, 
yielding  glass with slow  time evolution. 
With increasing 
the impurity size, the structural or translational  
disorder is also proliferated. 
\end{abstract}



\maketitle


Certain anisotropic particles such as KCN form 
a cubic crystal and, at  lower  temperatures, 
they undergo  an order-disorder phase transition, 
where the crystal structure changes 
 to  a noncubic one. 
Furthermore, with addition of impurities,  
the phase ordering   often 
occurs only on  small spatial scales, where  heterogeneous   
orientation fluctuations are pinned  \cite{ori}.  
In such systems 
softening of the shear modulus 
is observed, 
indicating direct coupling between 
the   molecular orientation and 
the acoustic phonons, and the 
molecules often have dipolar moments, yielding 
dielectric anomaly. 
These systems with frozen disorder 
have been identified  as 
orientational glass. As a similar example, 
metallic ferroelectric glass, called  relaxor,  
with frozen polar nanodomains 
have been studied extensively \cite{relax}.  
Recently,   a system of  
 off-stoichiometric intermetallic Ti-Ni 
was shown to be 
glassy martensite or strain glass,
exhibiting the shape-memory effect 
and the superelasticity  
 \cite{Ren}. 
 For a  one-component system of 
hard spheroids, 
Frenkel and Mulder \cite{Frenkel1} performed 
Monte Carlo simulation 
  to find isotropic liquid, nematic liquid, 
orientationally ordered solid, and 
orientationally disordered (plastic) solid. 
Theoretical approaches on strain glass  so far 
have been a phase-field theory 
with elastic field and 
a random temperature \cite{Saxena} 
and a spin-glass theory 
with elastic long-range interaction \cite{Lookman}.

In this Letter, we  propose   a simple microscopic   
model exhibiting  orientation-martensitic 
 phase transitions  
and  glass behavior. In two dimensions, 
we suppose elliptic  particles 
interacting  via an angle-dependent 
Lennard-Jones potential, 
 where their  positions are 
${\bi r}_i$ and their 
orientation vectors are 
${\bi n}_i=(\cos\theta_i,\sin\theta_i)$ ($i=1, \cdots, N$). 
There can be two particle 
species 1 and 2 
with  radii $\sigma_{1}$ and $\sigma_2$ 
and  numbers $N_1$ and $N_2$. We set  $N=N_1+N_2=4096$ 
  and change the composition $c=N_2/N$. 
The pair potential $U_{ij}$  between particles 
$i \in \alpha$ and $j\in \beta$ 
($\alpha,\beta=1,2$)  is 
 expressed as 
\be 
U_{ij}
=4\ep\bigg [(1+ 6A_{ij}) ({\sigma_{\alpha\beta}}/r_{ij})^{12}
-({\sigma_{\alpha\beta}}/r_{ij})^{6}\bigg ]  -C_{ij},
\en 
where  ${\bi r}_{ij}= 
{\bi r}_i -{\bi r}_j$, 
 $r_{ij}= |{\bi r}_{ij}|$,   
$\sigma_{\alpha\beta}=
(\sigma_\alpha + \sigma_\beta)/2$, and $\ep$ is 
the characteristic interaction energy. 
The particle anisotropy 
is taken into account by the angle factor,     
\be
A_{ij} = \chi_\alpha
({\bi n}_i\cdot\hrij)^2+\chi_\beta ({\bi n}_j\cdot\hrij)^2,
\en
where $\hrij=|\rij|^{-1} \rij$ represents the direction 
of $\rij$. We introduce the anisotropy parameters 
$\chi_1$ and $\chi_2$ for the two species.  
We truncate  the above potential 
for  $r_{ij}> r_c$  with 
the cut-off being $r_c=3\sigma_1$.  
We also set  
$C_{ij}=U_{ij}$  at $r_{ij}= r_c$ 
to ensure the continuity of the potential at $r_{ij}=r_c$, 
so there remains a  weak angle-dependence 
in   $C_{ij}$.  
Our potential is analogous to the Gay-Berne potential 
for anisotropic molecules \cite{Gay}, which has been used 
to simulate mesophases  of liquid crystals, 
and the Shintani-Tanaka potential 
with five-fold symmetry 
yielding frustrated particle configurations 
\cite{Shintani}.

\begin{figure}[htbp]
\begin{center}
\includegraphics[width=230pt,bb=0 0 183 346]{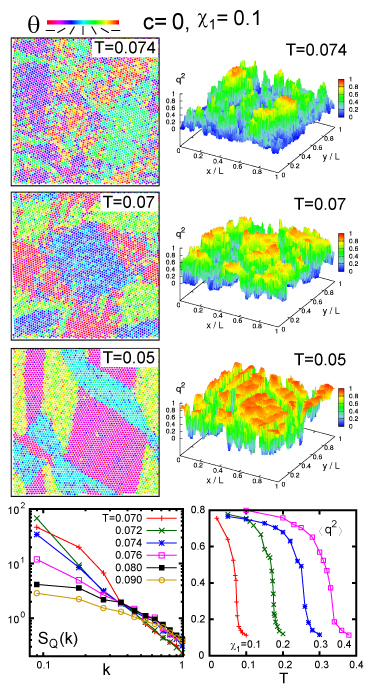}
\caption{Orientation angle $\theta_i$ (left) 
and order parameter amplitude $q^2_i$ (right) 
for $c=0$ and  $\chi_1=0.1$ at $T=0.074$, 
$0.07$, and $0.05$ from  above. Bottom left: 
Structure factor $S_Q(k)$ 
of the orientation fluctuations, 
growing for small $k$ in the range 
$T_2 \ls T \ls T_1$. Bottom right: 
Average amplitude   
$\av{q^2}=\sum_i q_i^2/N$ vs $T$ 
for $\chi_1= 0.1$, $0.2$, 
$0.3$, and $0.4$. }
\end{center}
\end{figure}

\begin{figure}[htbp]
\begin{center}
\includegraphics[width=250pt,bb=0 0 300 155]{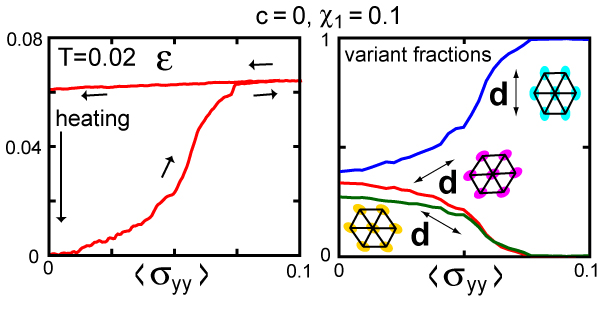}
\caption{Shape memory effect under uniaxial stretching 
along the $y$ axis at $T=0.02$ for $c=0$ and $\chi_1=0.1$. 
Left: Strain $\ve$ vs applied stress $\av{\sigma_{yy}}$ 
in units of $\epsilon/\sigma_1^2$. 
For  $\av{\sigma_{yy}}>0.075$, there remains only 
the variant  elongated along the $y$ axis.
After this cycle, the residual strain 
vanishes  upon heating to  $T=0.1$. 
Right: Fractions of the three variants during  the cycle, 
which are stretched along  the three crystal axes. 
}
\end{center}
\end{figure}

The total potential and kinetic 
energies   are $U= \sum_{i<j}U_{ij}$ and  
$K= \sum_i [m |d{\bi r}_i/dt|^2
+  I_\alpha |d\theta_i/dt|^2]/2$, respectively, 
where the two species have a common mass $m$   
and  inertia momenta $I_1$ and $I_2$. 
The Newton equations  of motions are 
\be
{m} 
\frac{d^2}{dt^2} 
{\bi r}_i=-\pp{U}{{\bi r}_i},\quad 
{I_\alpha} \frac{d^2}{dt^2} 
{\theta}_i= -\pp{U}{\theta_i}.
\en 
Since we treat 
equilibrium or  nearly steady states 
at a given temperature $T$, 
 we  attach  a Nos$\acute{\rm e}$-Hoover thermostat 
\cite{nose} to all the particles by adding the thermostat terms 
in Eq.(3). 
Space, time, and  $T$  will be  measured 
in  units of $\sigma_1$,  
 $\tau_0 = \sigma_1 \sqrt{m/\ep}$, and 
 $\epsilon/k_B$, respectively. 
In our simulation, we started  with 
a liquid  at  $T=2$, 
 quenched the system  to $T=0.35$   
below the melting temperature $T_m \sim 1.0$,   
and annealed it for $9000\tau_0$.  We then  lowered $T$ 
to a final  low temperature.

Assuming that the particles of the second species 
are  spherical and larger,    we set $\chi_2=0$,   
 $\sigma_2/\sigma_1=1.2$ or $1.4$, and   
  $c=0, 0.1$  or $0.2$. 
From Eq.(2)  the particles  of the species 1 
have  short and long 
diameters given by  $a_s= 2^{1/6}\sigma_{1}$ 
and $a_\ell=
 (1+12\chi_1 )^{1/6}a_s$, so their molecular area is $S_1= 
 \pi a_s a_\ell/4$ 
and their  inertia momentum 
is $I_1= (a_\ell^2+ a_s^2)  m  /4$, while 
 $I_2= \sigma_2^2  m  /2$ and $S_2= 
 \pi 2^{-5/3}\sigma_{2}^2$.
The packing fraction  $(N_1S_1+ N_2 S_2)/V$ 
is fixed at $0.95$ and 
the system length  is about $70\sigma_1$. 
For each particle $i$ 
of the first species, 
 we   introduce the orientation tensor ${\aQ}_i =
 \{ Q_{i\mu\nu}\}$ ($\mu,\nu=x,y$) as 
\bea
\aQ_i&=&
({1+N_{\rm b}^i})^{-1} ({\bi n}_i{\bi n}_i
 + \sum_{j\in {\rm bonded}} {\bi n}_j {\bi  n}_j) 
-\aI/2 \nonumber \\
&=&q_i ({\bi d}_i {\bi  d}_i -\aI/2),
\ena 
where $\aI$ is the unit tensor and 
${\bi d}_i$ is  the director with $|{\bi d}_i|=1$.  
The summation is 
 over the bonded particles $(|{\bi r}_{ij}| <3\sigma_1$) 
 of the first species  
with $N_{\rm b}^i$ being the number of 
these  bonded particles. When a  hexagonal lattice 
is formed, it   includes  
 the  second nearest neighbor particles.  
The  angle  of ${\bi d}_i$ 
varies  more smoothly than $\theta_i$. 
 The amplitude $q_i$ 
is  given by 
$q_i^2= 2 \sum_{\mu,\nu}  Q_{i\mu\nu}^2$.

First, we show numerical results  in  the one-component case 
($c=0$) with  $\chi_1=0.1$ to study  the 
orientation  phase transition 
on a hexagonal lattice. Here we use 
 the periodic  boundary condition at fixed volume, 
 but essentially the same results followed  
 at zero pressure.   In  Fig.1, we show  
the orientation angle $\theta_i$ 
of all the particles (left) 
and the order parameter amplitude $q^2_i$ (right)  
at $T=0.074, 0.07$, and $0.05$.  
From the angle snapshots we recognize 
emergence of three variants 
with  lowering  $T$ due to the underlying  hexagonal lattice. 
The left  bottom panel 
shows the structure factor 
$S_Q(k)= \av{ |Q_{2{\bi k}}|^2}$ 
for $Q_{2{\bi k}}= 
\sum_j (Q_{jxx}-Q_{jyy})
\exp({i{\bi k}\cdot{\bi r}_j})$, 
while the right  bottom panel displays 
the average $\av{q^2}= 
\sum_i q_i^2/N$ over all the particles 
for $\chi_1=0.1, 0.2,0.3$, and 0.4. 
The orientation  order develops 
gradually in a narrow region  
$T_2< T<T_1$, where 
$T_2\sim 0.070$ and $T_1\sim 0.076$ for $\chi_1=0.1$. 
The $T_1$ and $T_2$ increase with increasing $\chi_1$. 
In this temperature window,   
 a Berezinskii-Kosterlitz-Thouless  (BKT) phase 
\cite{Jose,Nelson} is  realized   
 between the low-temperature martensitic  phase 
 and the high-temperature orientationally disordered  phase, 
where  the orientation fluctuations 
are much enhanced at long wavelengths.
Though our system size 
 is still  small,  $S_Q(k)$ apparently grows as 
 $k^{\eta-2}$ for  $k\ls 0.5$, where   $\eta$   
 depends on $T$ (where $\eta \cong 0.05$ at $T=0.074$). 
We should note that  Bates and Frenkel \cite{Frenkel2} 
performed Monte Carlo simulation of 
two-dimensional rods to find the Kosterlitz-Thouless   phase 
transition.

For $T<T_2$, the three variants become distinct with sharp 
interfaces.  The surface tension 
between the variants is about $0.1 \epsilon/\sigma_1^2$ 
for $\chi_1=0.1$ (and 
is about  $0.2 \epsilon/\sigma_1^2$ 
for $\chi_1=0.2$).  
In  the pattern  at $T=0.05$ in Fig.1, 
the junction angles, at which two or more 
domain boundaries intersect,
are multiples of $\pi/6$. This geometrical 
constraint stops  the domain growth   at 
a characteristic size even without impurities \cite{Onukibook}. 
Similar   patterns were   observed 
on hexagonal planes in a number of 
experiments \cite{Kitano} 
and   were reproduced by 
 phase-field simulation \cite{Chen}.

In our model, the orientation order  induces 
 lattice deformations. 
As a result,  softening of the shear modulus $\mu$  
occurs  near the transition \cite{ori}, while 
the bulk modulus $K$ remains of order $20\epsilon/\sigma_1^2$.   
In fact, for $c=0$ and $\chi_1=0.1$, 
we have  $\mu \sim 3$ 
at $T=0.1$,  $\mu \sim 1.5$ 
at $T=0.08$, and $\mu \sim 5$ for  
$T \ls 0.05$  in units of $\epsilon/\sigma_1^2$.  
Each  variant at low $T$ 
is composed of  isosceles triangles elongated along 
one of the crystal axes,   
where side lengths are  $1.21 \sigma_1$ 
and  $1.11\sigma_1$ for $\chi_1=0.1$ 
at $T=0.05$ in Fig.1.  
 
In our system,  there arises 
 a  shape memory effect\cite{Ren}.  
 In Fig.2,  
we applied   a  stress $\av{\sigma_{yy}}$ 
along the $y$ axis   at  $T=0.02$  \cite{Rahman},  
treating  the surface along  it as  a free boundary.  
Initially,   the fractions of the 
three variants  were nearly close to $1/3$  and one 
 variant  was  elongated along the $y$ axis. 
 For very small $\ve<2\times 10^{-3}$, 
the system deformed elastically with $\mu \sim 2$. 
However, for  $2\times 10^{-3}<\ve< 0.075$, 
 the fraction of the favored variant increased 
 up to unity, while 
  those of the disfavored ones decreased. This inter-variant 
transformation occurs    without   
defect formation. In the next step   
$\av{\sigma_{yy}}$ was decreased slowly from 
$0.1\epsilon/\sigma_1^2$. 
On this return path, 
 the solid was composed of 
 the favored variant only.   
At vanishing stress, 
   there remained a  remnant strain, but it    
disappeared  upon  heating to $T=0.1$ 
 above the transition.  Here, we may 
 define the effective shear modulus $\mu_e$ 
by $[\p \epsilon/\p \av{\sigma_{yy}}]^{-1}= 
4K\mu_e/(K+\mu_e) \cong 4\mu_e$. 
Then $\mu_e \sim 0.1-0.8$ 
during the inter-variant transformation 
and $\mu_e\sim 5$ on the return path. 

\begin{figure}[htbp]
\begin{center}
\includegraphics[width=230pt,bb=0 0 182 276]{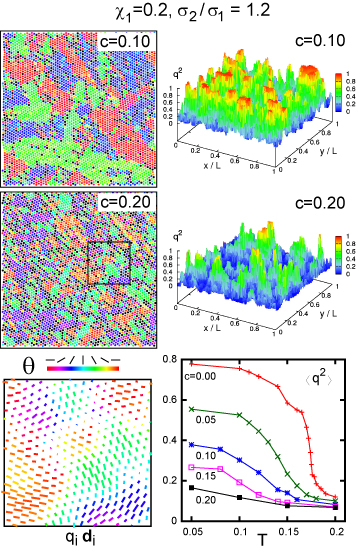}
\caption{Frozen patterns of  angle $\theta_i$ (left) 
and order parameter amplitude $q_i$ (right) with impurities 
for $c=0.1$ and 0.2, where 
$\chi_1=0.2$ at $T=0.05$.  
Bottom left: Expanded 
snapshot of  $q_i {\bi d}_i$ 
in a box in the upper panel, showing pinned 
mesoscopic  order or strain. 
Bottom right: $\av{q^2}$ vs $T$ for various $c$. 
}
\end{center}
\end{figure}

\begin{figure}[htbp]
\begin{center}
\includegraphics[width=210pt,bb=0 0 158 97]{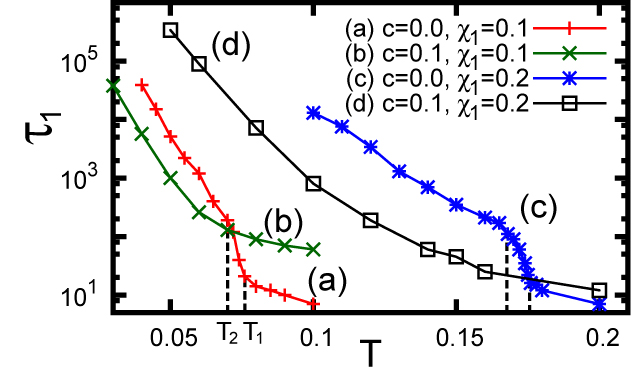}
\caption{Orientation relaxation time $\tau_1$ 
from the time-correlation function $G_1(t)$ 
for (a) $c=0$ and $\chi_1=0.1$,  (b) 
$c=0.1$ and $\chi_1=0.1$, 
 (c) $c=0$ and $\chi_1=0.2$, and  
 (d) $c=0.1$ and $\chi_1=0.2$.
It represents the turnover time. 
For (a) and (c), 
 $\tau_1$ grows steeply in the 
Berezinskii-Kosterlitz-Thouless  phase ($T_2<T<T_1$).  
For strain glass  (b) and (d), this phase is 
nonexistent and $\tau_1$ 
grows as $T$ is lowered.  
 }
\end{center}
\end{figure}

Next, in Fig.3,  we present  examples of  strain glass  
with  impurities, where  
$\sigma_2/\sigma_1=1.2$ and $\chi_1=0.1$. 
For $c=0.1$ and 0.2,    
 there appeared a few tens of particles  
with coordination numbers different from six 
 in a single crystal. 
In our model, the  elliptic 
particles tend to be parallel to 
the surface of the larger spherical ones, 
resulting in anchoring  of the orientation. 
We can see that the size of the  domains 
 decreases with increasing $c$, where  
the impurities suppress the 
development of the orientation 
order. In addition, the BKT phase 
disappeared in  these examples. 
We  also observed a shape-memory effect 
even in orientational 
glass states \cite{Ren}, where small disfavored 
domains  were replaced by the favored 
ones  upon stretching.

\begin{figure}[htbp]
\begin{center}
\includegraphics[width=230pt,bb=0 0 168 109]{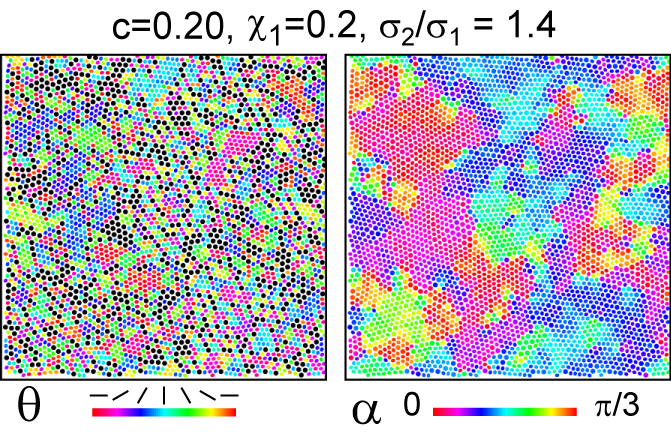}
\caption{
Orientation angle $\theta_i$ (left) 
and  six-fold bond orientation 
angle $\alpha_i$ in Eq.(7) (right) in polycrystal 
for    $\sigma_2/\sigma_1=1.4$ 
and  $c=0.2$,  
where $\chi_1=0.2$   and $T=0.05$. 
}
\end{center}
\end{figure}

Now, we discuss the dynamics. 
Let us consider the 
time-dependent angle-distribution function,  
\be 
G(t,\varphi)
=\sum_j \av{\delta(
\theta_j(t+t_0)-\theta_j(t_0)-\varphi)}/N_1 ,  
\en 
where the average $\av{\cdots}$ is taken over 
the initial time $t_0$ and over several runs. 
We are interested in the first two 
moments $G_1(t)$ and $G_2(t)$. For $p=1, 2,\cdots$, we define  
\be 
G_p(t)= \int_0^{2\pi} d\varphi  
G(t,\varphi) \cos(p\varphi) /2\pi, 
\en 
which decays from unity 
on a  time scale of $\tau_p$. 
Here, $\tau_1$ is the inverse frequency of 
the turnover motions    $\theta_j(t)\to  
\theta_j(t)\pm \pi$ at low $T$, 
while $\tau_2$ is  the randomization time  of 
$\cos({2\theta_j(t)-2\theta_j(0)})$. 
Each turnover motion takes place quickly. 
We find that  $G(t,\varphi)$ 
exhibits a peak of   the   form  
$A(t)\exp[-(\varphi-\pi)^2/2\sigma^2]/\sqrt{2\pi}\sigma$ 
for small $\varphi -\pi$ at low $T$, where $\sigma \sim 0.45$.  
The coefficient  $A(t)$ grows linearly as $  0.5t/\tau_1$ 
for   $t\ll  \tau_1$ 
and  tends 
to a constant ($\cong 1/2$) for $t\gg \tau_1$.
The fitting 
$G_1(t) = 
\exp[-(t/\tau_1)^\beta]$ 
fairly holds, 
 where  $\beta$ 
 decreases from unity 
 to about 0.5 as $T$ is lowered.

For $c=0$, 
 $\tau_1$ increases steeply 
in the BKT phase 
and  $\tau_1^{-1} \propto 
\exp(-T_0/T)$ for $T \ls T_2$, where  
(a) $T_0 \sim 0.40$ at  
$\chi_1=0.1$ 
and (c) $T_0 \sim 1.2$ at  $\chi_1=0.2$ in Fig.4. 
In addition, 
 $\tau_1\sim \tau_2$  
for $T \gs T_1$ but $\tau_2/\tau_1 \gg 1$   
for $T\ls T_2$. 
In fact, for $\chi_1=0.1$,   
 the ratio is about $ 10^2$ 
 at  $T= 0.07$ 
and is about $10^3$ at  $T= 0.06$.  
On the other hand, in glassy 
states with  impurities, the relaxation 
behavior is more complicated due to 
the pinning effect, but 
the turnover motions 
still occur and $\tau_2\gg \tau_1$ holds. 
 Figure 4  shows  that    $\tau_1$ for $c=0.1$ 
is longer in the disordered phase but 
is shorter in the ordered phase 
than in the pure system.

For $\sigma_2/\sigma_1=1.2$ 
and for $c=0.1$ and 0.2, 
the crystal structure is    
little affected by the orientation 
fluctuations. For a  larger size ratio, 
the  structural or positional 
disorder is more enhanced, 
eventually resulting in 
 polycrystal  and glass \cite{Hama}.   In Fig.5,  
we realize a polycrystal state for  
$\sigma_2/\sigma_1=1.4$, $\chi_1=0.2$, and $c=0.2$, 
where black points represent the impurities. 
The left panel displays  $\theta_j$, 
where  there remains noticeable orientation order with 
$\av{q^2}=0.2$.  The right panel displays  
 the positional  sixfold orientation 
angle $\alpha_j$ \cite{Nelson,Hama}. 
Here, for each elliptic particle $j$,  
we define $\alpha_j$ in the range 
$0\le \alpha_j <\pi/3$ by 
\be
\sum_{k\in\textrm{\scriptsize{bonded}}} \exp [6i\theta_{jk}]
=Z_j \exp[{6i\alpha_j}], 
\en 
where $\theta_{jk}$ 
is the angle of ${\bi r}_{jk}= 
{\bi r}_k-{\bi r}_j$ with respect to the $x$ axis. 
We  set  $|{\bi r}_{jk}| <1.7\sigma_1$ and $Z_j>0$.

In summary, we have presented an  angle-dependent 
Lennard-Jones potential  to simulate orientation or 
martensitic transitions. 
We have  added impurities,   
which pin orientation 
and strain fluctuations on mesoscopic scales. 
In future, we should examine the impurity pinning  
on the glass transition in detail 
by systematically changing the composition 
and the size ratio \cite{Hama}. 
Competition of the orientational 
and translational glass behaviors 
should also be  studied.  
We will shortly report 
 three-dimensional 
simulation results, where 
inclusion of  the dipolar 
interaction will  enrich the problem.

\end{document}